\documentclass[useAMS,usenatbib]{mn2e}

\usepackage{bm}
\usepackage{graphicx}

\DeclareMathSymbol{\varOmega}{\mathord}{letters}{"0A}
\DeclareMathSymbol{\varSigma}{\mathord}{letters}{"06}
\DeclareMathSymbol{\varPsi}{\mathord}{letters}{"09}

\newcommand{\vc}[1]{\bm{#1}}
\newcommand{\de}{\mathrm{d}}
\newcommand{\dpa}{\partial}

\newcommand{\nab}{\vc{\nabla}}

\newcommand{\apj}{ApJ}
\newcommand{\apjl}{ApJL}
\newcommand{\aap}{A\&A}
\newcommand{\mnras}{MNRAS}

\newcommand{\Eq}[1]{equation (\ref{#1})}

\newcommand{\Fig}[1]{Fig.~\ref{#1}}

\newcommand{\Tab}[1]{Table \ref{#1}}

\title[Turbulent diffusion in protoplanetary discs]
{Turbulent diffusion in protoplanetary discs:\\ The effect of an imposed magnetic field}
\author[A. Johansen et al.]
    {A. Johansen$^1$, H. Klahr$^1$ and A.J. Mee$^2$ \\
    $^1$Max-Planck-Institut f\"ur Astronomie, K\"onigstuhl 17, 69117
    Heidelberg, Germany\\
    $^2$School of Mathematics and Statistics, University of Newcastle upon Tyne,
    NE1 7RU, UK}
\begin{document}
\date{Accepted 200- December --. Received 200- December --; in original form 200- October --}
\pagerange{\pageref{firstpage}--\pageref{lastpage}} \pubyear{2006}
\maketitle
\label{firstpage}
\begin{abstract}
  We study the effect of an imposed vertical magnetic field on the turbulent
  mass diffusion properties of magnetorotational turbulence in protoplanetary
  discs.  It is well-known that the effective viscosity generated by the
  turbulence depends strongly on the magnitude of such an external field. In
  this letter we show that the turbulent diffusion of the flow also grows, but
  that the diffusion coefficient does not rise with increasing vertical field
  as fast as the viscosity does. The vertical Schmidt number, i.e. the ratio
  between viscosity and vertical diffusion, can be close to $20$ for high field
  magnitudes, whereas the radial Schmidt number is increased from below unity
  to around $3.5$. Our results may have consequences for the interpretation of
  observations of dust in protoplanetary discs and for chemical evolution
  modelling of these discs.
\end{abstract}
\begin{keywords}
  accretion, accretion discs ---
  diffusion ---
  MHD ---
  turbulence ---
  planetary systems: formation ---
  planetary systems: protoplanetary discs
\end{keywords}
\section{Introduction}

Planets form out of micrometer-sized dust grains that are embedded in the gas
in protoplanetary discs \citep[see][ for a recent review]{Dominik+etal2006}.
The observed infrared radiation from protoplanetary discs comes primarily from
micron-sized grains, although observations at longer wavelengths show that some
discs have large populations of grains with sizes up to mms and cms
\citep[e.g.][]{Rodmann+etal2006}. Turbulent motions in the gas play a big role
in the dynamics of chemical species and solids, at least as long as the solids
are smaller than a few ten meters. Thus an understanding of how dust grains and
chemical species move under the influence of turbulence is vital for our
understanding of the physical processes that take place in protoplanetary discs
and the observational consequences
\citep{Ilgner+etal2004,IlgnerNelson2006,Willacy+etal2006,Dullemond+etal2006,
Semenov+etal2006}.

Turbulence has a number of effects on the embedded dust grains. Larger grains
(rocks and boulders) can be trapped in the turbulent flow due to their marginal
coupling to the gas \citep{BargeSommeria1995}, whereas smaller grains feel the
effect of the turbulence as a combination of diffusion and simple advection.
Any bulk motion of the gas, e.g.\ turbulent motion with a turn-over time that
is longer than the time-scale that is considered or even a radial accretion flow,
leads to an advective transport of the dust grains rather than diffusion. The
turbulent transport acts as diffusion only when the considered time-scale is
longer than the eddy turn-over time.
The turbulent diffusion coefficient of the grains, $D_{\rm t}=\delta c_{\rm
s}^2 \varOmega_0^{-1}$, is
often assumed to be equal to the turbulent viscosity of the gas flow
$\nu_{\rm t}=\alpha c_{\rm s}^2 \varOmega_0^{-1}$. Here a
non-dimensionalisation with sound speed $c_{\rm s}$ and Keplerian frequency
$\varOmega_0$ is used \citep{ShakuraSunyaev1973}.
The Schmidt number, a measure of the relative strength of turbulent viscosity
and turbulent diffusion, is defined as the ratio ${\rm Sc} = \nu_{\rm
t}/D_{\rm t} = \alpha/\delta$. Several recent works have
measured the turbulent diffusion coefficient directly from numerical
simulations of magnetorotational turbulence \citep{BalbusHawley1991}.
The
simulations by Johansen \& Klahr (2005, hereafter JK05) yielded a Schmidt
number that is around unity for radial diffusion, whereas Carballido, Stone, \&
Pringle (2005, hereafter CSP05) found a value as high as $10$. The vertical
Schmidt number, measured both by JK05, \citet{Turner+etal2006} and by
\citet{FromangPapaloizou2006}, gives more consistently a number between 1 and
3. Here it is worthy of note that \citet{Turner+etal2006} consider stratified
discs, and \citet{FromangPapaloizou2006} even include the effect of a
magnetically dead zone without turbulence around the mid-plane
\citep{Gammie1996,FlemingStone2003}.

This letter addresses the discrepancy between the diffusion properties of
turbulence in protoplanetary discs reported in the literature. We show that a
vertical imposed magnetic field affects the diffusion coefficient strongly. It
is known that a net vertical field component leads to turbulence with a
stronger angular momentum transport \citep{Hawley+etal1995}. We perform computer simulations of
magnetorotational turbulence for various values of the vertical field and find
that turbulent diffusion does not increase as much as the viscosity increases.
Thus the ratio between viscous stress and diffusivity, i.e.\ the Schmidt
number, also increases with the magnitude of the external field. As a result we are able
to give a possible explanation for the discrepancy in the radial Schmidt
numbers found in the literature.

\section{Sources of an external magnetic field}

The properties of any external magnetic field threading protoplanetary discs
are not well-known. Close to the central object there is an interaction with
the possibly dipolar or maybe quadrupolar magnetic field of the young stellar
object. Also the occurrence of jet phenomena indicates that at least for the
originating zone of the jet, e.g. a few protostar radii, there should be a
large scale vertical magnetic field
\citep[e.g.][]{FendtElstner1999,Vlemmings+etal2006}. However, at larger orbital
distances relevant for planet formation, it is not obvious what the global
field configuration should look like.

To get some physical insight into the role of an external magnetic field in the
dynamics of protoplanetary discs, we do here some rough estimations for two
cases, either that the field originates in the central object, or that it comes
from the molecular cloud core out of which the disc formed.

\subsection{Protostar}

The dipolar field of the central protostar dominates the gas pressure of the
disc until a certain inner disc radius $R_{\rm in}$. This is typically a few
times the protostellar radius \citep{Camenzind1990,Koenigl1991,Shu+etal1994}.
Beyond $R_{\rm in}$ the interaction between the dipole field and the accretion
disc is strongly unstable and leads to an opening up of the protostellar dipole
field lines \citep{MillerStone1997,FendtElstner2000,Kueker+etal2003}. Even if
the protostar could retain its dipolar field at larger orbital radii, the
magnetic pressure exerted by the field lines would fall so quickly with orbital
radius [$B_z^2(r) \propto r^{-6}$] that it would be completely unimportant at
several AU from the protostar where the gas planets are believed to form.

\subsection{Molecular cloud}

In molecular cloud cores the magnetic field, $B_{\rm cloud}$, can be as large
as $\sim 100\,\umu{\rm G}$ \citep{Bourke+etal2001}. The gas pressure in the
disc can be written as $P=c_{\rm s}^2 \rho$, where $c_{\rm s}$ is the sound
speed and $\rho$ is the gas density.  The mid-plane density of an exponentially
stratified disc with scale height $H$ depends on the column density $\varSigma$
as $\rho=\varSigma/(\sqrt{2\pi}H)$.  The scale-height to radius ratio $H/r$,
which also corresponds to the ratio of local sound speed to Keplerian speed
$v_{\rm K}$, can be used to rewrite the gas pressure at the mid-plane of the
disc as,
\begin{equation}
  P = \left( \frac{H}{r} \right)^2 v_{\rm K}^2
      \frac{\varSigma(r)}{\sqrt{2 \pi} (H/r) r}
    = \frac{H}{r} \frac{G M_\star}{\sqrt{2 \pi}} \frac{\varSigma(r)}{r^2} \, .
  \label{eq:Pgas}
\end{equation}
The plasma beta of the external magnetic field is defined as the ratio
between gas pressure and magnetic pressure $\beta = P/P_{\rm mag}$.  One can
write the following scaling for the plasma beta $\beta_{\rm cloud}$ due to the
magnetic field from the molecular cloud,
\begin{eqnarray}
  \beta_{\rm cloud} &=& 5.9\cdot10^7
      \left( \frac{H/r}{0.1} \right)
      \left( \frac{M_\star}{M_\odot} \right) 
      \left( \frac{B_{\rm cloud}}{\umu{\rm G}} \right)^{-2} \nonumber \\
      & & \hspace{2.0cm}
      \left( \frac{\varSigma}{1\,{\rm g\,cm^{-2}}}\right)
      \left( \frac{r}{{\rm 100 AU}} \right)^{-2} \, .
\end{eqnarray}
Here $\beta_{\rm cloud}$ has a falling trend with $r$ because the low gas
density in the outer part of the disc makes the magnetic pressure more
important there. For a sufficiently strong cloud field, the plasma beta could
be relatively low at a disc radius of several hundred astronomical units.
\begin{table*}
  {\scriptsize
  \centering
  \begin{minipage}{140mm}
  \caption{Measured turbulent viscosity and diffusion coefficients}
  \label{t:runs}
  \begin{tabular}{rrrrrrrrrrrrrr}
    \hline
    Run & $L_x$$\times$$L_y$$\times$$L_z$ & $B_0$ & $\beta$
        & $\alpha$ & ${\rm Ma}_x$ & ${\rm Ma}_y$ & ${\rm Ma}_z$
        & $\delta_x$ & ${\rm Sc}_x$ & $\delta_z$ & ${\rm Sc}_z$ \\
    \hline
    A & $1.32$$\times$$1.32$$\times$$1.32$ & $0.00$ & $\infty$
      & $0.0028\pm0.0004$ & $0.053$ & $0.053$ & $0.041$
      & $0.0031$ & $0.90$ & $0.0016$ & $1.75$ \\
    B & $-$ & $0.01$ & $20000$
      & $0.0078\pm0.0015$ & $0.079$ & $0.092$ & $0.064$
      & $0.0058$ & $1.34$ & $0.0031$ &  $2.52$ \\
    C & $-$ & $0.03$ & $ 2222$
      & $0.0367\pm0.0142$ & $0.197$ & $0.185$ & $0.140$
      & $0.0225$ & $1.63$ & $0.0092$ & $3.99$ \\
    D & $-$ & $0.05$ & $  800$
      & $0.1811\pm0.0773$ & $0.416$ & $0.300$ & $0.181$
      & $0.0574$ & $3.16$ & $0.0123$ & $14.72$ \\
    E & $-$ & $0.07$ & $  408$
      & $0.5529\pm0.0964$ & $0.761$ & $0.421$ & $0.330$
      & $0.1984$ & $2.79$ & $0.0300$ & $18.43$ \\
    A4 & $1.00$$\times$$4.00$$\times$$1.00$ & $0.00$ & $\infty$
      & $0.0015\pm0.0002$ & $0.055$ & $0.036$ & $0.031$
      & $0.0017$ & $0.88$ & $0.0009$ & $1.71$ \\
    B4 & $-$ & $0.01$ & $20000$
      & $0.0038\pm0.0009$ & $0.079$ & $0.057$ & $0.052$
      & $0.0038$ & $1.00$ & $0.0024$ &  $1.58$ \\
    C4 & $-$ & $0.03$ & $ 2222$
      & $0.0414\pm0.0176$ & $0.206$ & $0.182$ & $0.134$
      & $0.0177$ & $2.34$ & $0.0078$ & $5.31$ \\
    D4 & $-$ & $0.05$ & $  800$
      & $0.0793\pm0.0371$ & $0.279$ & $0.239$ & $0.179$
      & $0.0268$ & $2.96$ & $0.0091$ & $ 8.71$ \\
    E4 & $-$ & $0.07$ & $  408$
      & $0.1242\pm0.0694$ & $0.366$ & $0.291$ & $0.221$
      & $0.0356$ & $3.49$ & $0.0121$ & $10.26$ \\
    \hline
  \end{tabular}
  \end{minipage}
  }
\end{table*}

\section{Simulations}

We simulate a protoplanetary disc in the shearing sheet approximation
\citep[e.g.][]{GoldreichTremaine1978,Brandenburg+etal1995,Hawley+etal1995}.
Here a local coordinate frame corotating with the disc with the Keplerian
rotation frequency $\varOmega_0$ at a distance $r_0$ from the central source of
gravity is considered. The coordinate system is oriented so that $x$ points
radially away from the central object, $y$ points in the azimuthal direction
parallel to the the Keplerian flow, and $z$ points normal to the disc along the
Keplerian rotation vector $\vc{\varOmega}_0$. Numerical calculations are
performed using the Pencil Code \citep[a finite difference code that uses sixth
order symmetric space derivatives and a third order time-stepping scheme,
see][]{Brandenburg2003}.

\subsection{Gas}

Considering the velocity field $\vc{u}$ relative to the Keplerian flow
$u_y^{(0)}=-(3/2) \varOmega_0 x$, the equation of motion of the gas is
\begin{eqnarray}
  &&\frac{\dpa \vc{u}}{\dpa t} + (\vc{u}\cdot\nab)\vc{u} + u_y^{(0)}
  \frac{\dpa \vc{u}}{\dpa y} = \vc{f}(\vc{u}) -
  c_{\rm s}^2 \nab \ln \rho \nonumber \\
  & & \hspace{2.8cm} + \frac{1}{\rho} \vc{J} \times (\vc{B}+B_0\hat{\vc{z}}) +
  \vc{f}_{\rm \nu} \, .
  \label{eq:gasmomentumeq}
\end{eqnarray}
The left-hand-side of \Eq{eq:gasmomentumeq} contains terms for both the
advection by the velocity relative to the Keplerian flow and for the advection
by the Keplerian flow itself. The terms on the right-hand side are the modified
Coriolis force, 
\begin{equation}
  \vc{f}(\vc{u}) = \pmatrix{2 \varOmega_0 u_y \cr -
  \frac{1}{2} \varOmega_0 u_x \cr 0} \, ,
  \label{eq:fCor}
\end{equation}
which takes into account that the Keplerian velocity profile is advected with
any radial motion, the force due to the isothermal pressure gradient with a
constant sound speed $c_{\rm s}$, the Lorentz force (including the contribution
from an imposed vertical field of strength $B_0$) and the viscous force
$\vc{f}_\nu$ that is used to stabilise the numerical scheme. The viscosity term
is a combination of sixth order hyperviscosity and a localised shock capturing
viscosity. The use of hyperviscosity, hyperdiffusion and hyperresistivity is
explained in JK05. For the shock viscosity, where extra bulk viscosity is added
in regions of flow convergence, we refer to \cite{Haugen+etal2004} for a
detailed description.

The evolution of the mass density is solved for in the continuity equation
\begin{equation}
  \frac{\dpa \rho}{\dpa t} + \vc{u} \cdot \nab \rho + u_y^{(0)} \frac{\dpa \rho}{\dpa y} =
  - \rho \nab \cdot \vc{u} + f_{\rm D} \, ,
  \label{eq:gasconteq}
\end{equation}
where $f_{\rm D}$ is a combination of sixth order hyperdiffusion and shock
diffusion. The magnetic field evolves by the induction equation which we write
in terms of the magnetic vector potential $\vc{A}$,
\begin{equation}
  \frac{\dpa \vc{A}}{\dpa t} + u_y^{(0)} \frac{\dpa\vc{A}}{\dpa y} =
  \frac{3}{2} \varOmega_0 A_y \hat{\vc{x}} + \vc{u} \times
  (\vc{B}+B_0\hat{\vc{z}}) +
  \vc{f}_\eta \, .
  \label{eq:inductioneq}
\end{equation}
Again we use sixth order hyperresistivity and shock resistivity, through the
function $\vc{f}_\eta$, in regions of strong flow convergence. The value of
$B_0$ sets the strength of an external vertical magnetic field.

\subsection{Dust particles}

The turbulent diffusion coefficient $D_{\rm t}$ of the flow is measured by
letting dust grains settle to the mid-plane of the turbulent disc. The dust
layer is represented as individual particles each with a
position $\vc{x}^{(i)}$ and velocity vector $\vc{v}^{(i)}$ (measured relative
to the Keplerian velocity $u_y^{(0)}\hat{\vc{y}}$). The gas acts on a
dust particle through a drag force that is proportional to but in the opposite
direction of the difference between the velocity of the particle and the local
gas velocity. The dust grains do not interact mutually and do not have any
feedback on the gas. The equation of motion of the dust particles is
\begin{equation}
  \frac{\de \vc{v}^{(i)}}{\de t} = \vc{f}(\vc{v}^{(i)})
      - \frac{1}{\tau_{\rm f}} \left(\vc{v}^{(i)}-\vc{u}\right) + \vc{g}\, ,
  \label{eq:eqmotp}\\
\end{equation}
where the modified Coriolis force $\vc{f}$ is defined in \Eq{eq:fCor},
$\tau_{\rm f}$ is the friction time and $\vc{g}$ is an imposed
gravitational field (see below).  We assume in the following that $\tau_{\rm
f}$ is constant and thus independent of the relative velocity between the grain
and the surrounding gas. In protoplanetary discs this is a valid assumption for
sufficiently small dust grains \citep{Weidenschilling1977}. We use a value of
$\varOmega_0 \tau_{\rm f}=0.01$ which is small enough that the diffusion
coefficient should not differ significantly from that of a passive scalar
(which can be seen as a dust grain in the limit of a vanishingly small friction
time).  This value is also large enough that the computational time-step is set
by the Courant criterion for the gas and not by the friction force in the dust
equations.

The particles change positions according to the dynamical equation
\begin{equation}
  \frac{\de \vc{x}^{(i)}}{\de t} = \vc{v}^{(i)} + u_y^{(0)} \hat{\vc{y}} \, .
\end{equation}
Under the effect of a special gravity field acting on the dust only, $\vc{g}$
in \Eq{eq:eqmotp}, the particles fall either to the horizontal mid-plane of the
disc, in the case of a vertical gravity field $\vc{g}=g_z(z)\hat{\vc{z}}$, or
to a vertical ``mid-plane'' in the case of a radial gravity field
$\vc{g}=g_x(x)\hat{\vc{x}}$.  We use a sinusoidal expression $g_i = -g_0
\sin(k_i x_i)$ with a wavelength that is equal to the size of the simulation
box.  In the equilibrium state, the sedimentation is balanced by the turbulent
diffusion away from the mid-plane, and the dust number density $n$, for the
case of a vertical gravity field, is given by (see JK05)
\begin{equation}
  \ln n(z) = \ln n_1 + \frac{\tau_{\rm f} g_0}{k_z D_z^{\rm (t)}} \cos(k_z z) \, ,
  \label{eq:lnnz_equi}
\end{equation} 
where $n_1$ is an integration constant. The equivalent expression for the
radial gravity case is found simply by replacing $z$ by $x$ in
\Eq{eq:lnnz_equi}.

We run simulations with different values of the external magnetic field
strength $B_0$ between $0$ and $0.07$, corresponding to a $\beta$ ranging from
infinity down to approximately $400$. Our computational unit of velocity is the
constant sound speed $c_{\rm s}$, length is in units of the disc scale-height
$H$, and density is measured in units of mean gas density $\rho_0$. In these
units the turbulent viscosity and the turbulent diffusion coefficient,
$\nu_{\rm t}$ and $D_{\rm t}$, are numerically equal to the dimensionless
coefficients $\alpha$ and $\delta$.  The unit of the magnetic field is then
$[B]=c_{\rm s} \sqrt{\mu_0 \rho_0}$ and is chosen such that $\mu_0=1$.  For
each value of $B_0$ we run one simulation with a vertical and one simulation
with a radial gravitational field on the dust particles. The diffusion
coefficients $\delta_x$ and $\delta_z$ are found by fitting a cosine
function to the logarithmic dust density.  From the amplitude we then determine
the diffusion coefficient using \Eq{eq:lnnz_equi}. The run parameters and the
results are shown in \Tab{t:runs}. Two simulation box sizes are considered, a
square box with a side length of $1.32$ and an elongated box with
$(L_x,L_y,L_z)=(1.0,4.0,1.0)$ \citep[similar to the setup of][]{Sano+etal2004}. For the first case
we use a resolution of $64^3$ grid points and
1,000,000 dust particles. Simulations
with $128^3$ grid points were done by JK05 and showed only small differences
from the $64^3$ simulations in the measured Schmidt numbers.  Each model is run
for twenty local orbits, i.e.  $20 \times 2\pi\varOmega_0^{-1}$, of the disc.
The runs with an elongated box are done with $64\times256\times64$ grid
points and 4,000,000 dust particles.

\section{Results}

\begin{figure}
  \includegraphics[width=\linewidth]{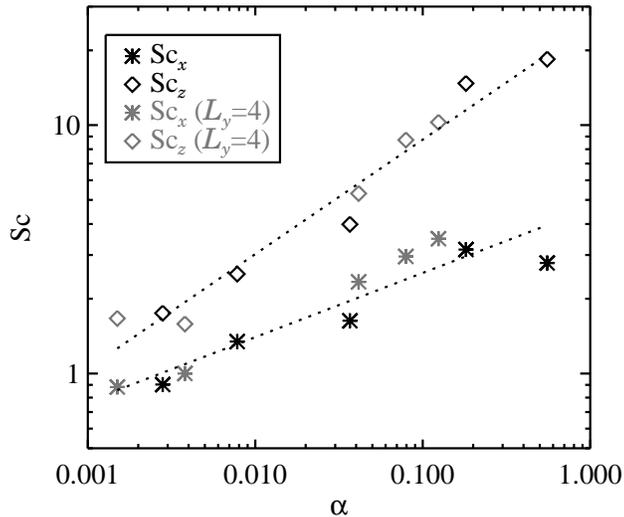}
  \caption{The Schmidt number plotted against the $\alpha$ value and the best
      power-law fit (dotted lines). The best fit has ${\rm Sc}_x=4.6
      \alpha^{0.26}$ and ${\rm Sc}_z = 25.3 \alpha^{0.46}$.}
  \label{f:Sc_alpha}
\end{figure}
For each value of the imposed magnetic field we have measured the
$\alpha$-value from the Reynolds and Maxwell stress tensors (see
\Tab{t:runs}).  The $\alpha$-value grows approximately exponentially with $B_0$. An
$\alpha$-value close to unity can be reached already for $B_0=0.07$
(corresponding to $\beta\simeq400$). A similar investigation into the
dependence of $\alpha$ on an imposed vertical field was undertaken by
\citet{Hawley+etal1995}. Comparing with Table 1 in that work, one sees that
there is a relatively good agreement between those results and ours.
Magnetorotational instability with an imposed vertical field develops into a
``channel'' solution
\citep{HawleyBalbus1992,GoodmanXu1994,SteinackerHenning2001}, characterised by
the transfer of the most unstable MRI mode to the the largest scale of the
simulation box
and the subsequent decay of this large scale mode \citep{SanoInutsuka2001}.
Sufficiently strong vertical fields can even cause stratified discs to break up
altogether \citep{MillerStone2000}. The creation and destruction of the
unstable channel solution gives significant temporal fluctuations in the
measured stresses, evident in the standard deviation of the turbulent viscosity
in \Tab{t:runs} \citep[see also Fig. 1 of][]{SanoInutsuka2001}.

For measuring the turbulent diffusion coefficient we consider the logarithmic
number density of the dust particles averaged from $10$ to $20$ orbits. We have
chosen to calculate the diffusion coefficient directly from this average state,
rather than calculating it from the instantaneous dust density at a given time
$t$, because large-scale advection flow only works as diffusion when averaged
over sufficiently long times. The average dust density was found to be in
excellent agreement with the cosine distribution of \Eq{eq:lnnz_equi} with a
deviation from a perfect cosine of less than $5\%$ for all simulations. Thus
diffusion is a good description of the turbulent transport over long
time-scales. This is partly due to the fact that we consider diffusion at the
largest scale of the flow, i.e.\ at a scale that is similar to or larger than
the energy injection scale of the MRI. Diffusion over length scales that are
smaller than the energy injection scale should be weaker, because dust density
concentrations at small length scales are not stretched by the full velocity
amplitude of the larger scales, but only by the velocity difference that the
larger scales exert over the much narrower dust concentration. The $\exp(\cos)$
equilibrium state for the dust density,
however, has almost all of its power at the largest scale of the simulation box,
so any scale-dependency of the diffusion coefficient should not have any
influence on the equilibrium state (the fact that the logarithmic dust density
in the equilibrium state is a cosine supports this).

The measured turbulent diffusion coefficients are written in
\Tab{t:runs}. It is evident that the turbulent diffusion coefficient does not
increase as fast with increasing vertical field as the turbulent viscosity
does. In \Fig{f:Sc_alpha} we plot the vertical and radial Schmidt numbers as a
function of $\alpha$. Both 
Schmidt numbers approximately follow a power law with $\alpha$.  Making a
best-fit power law, we find the empirical connections
\begin{eqnarray}
  {\rm Sc}_x &=& \,\,\, 4.6 \alpha^{0.26} \, , \\
  {\rm Sc}_z &=& 25.3 \alpha^{0.46} \, .
\end{eqnarray}
Considering the two box sizes individually (black and grey symbols in
\Fig{f:Sc_alpha}), the radial Schmidt number is seen to rise slightly faster
with increasing $\alpha$ in the case of the elongated box with $L_y=4$, whereas the
vertical Schmidt number follows a trend that is independent of the box size.
In ideal MHD simulations with $\beta=400$, CSP05 find a radial Schmidt number
of around $10$. Using a similar value for $\beta$, we find that the radial
Schmidt number rises from unity in the case of no external field to $\sim3-4$
when $\beta\simeq400$. This may explain at least part of the discrepancy
between the results by CSP05 and JK05. The box size used in CSP05 is
$1.0\times6.28\times1.0$, and is thus comparable to our elongated box. We have
tried with $L_y=6.28$ as well, but found no significant difference in the
results.
\begin{figure}
  \includegraphics[width=\linewidth]{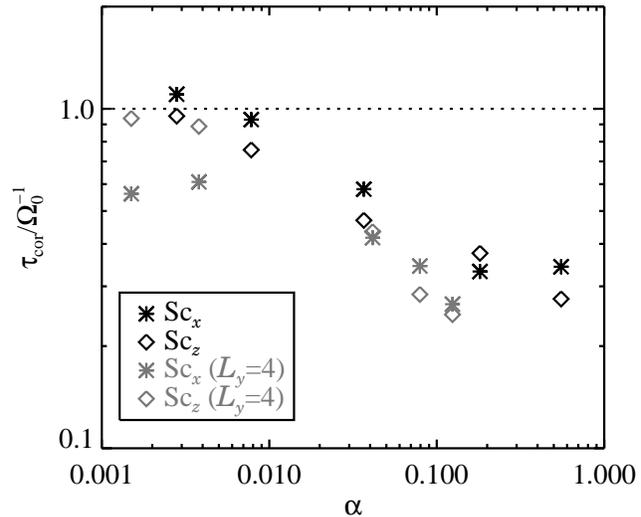}
  \caption{The correlation time of the turbulent mixing coefficients versus the
      $\alpha$-value. The correlation times fall significantly with increasing
      $\alpha$.}
  \label{f:taucor_alpha}
\end{figure}

It is interesting to note that \citet{FromangPapaloizou2006} have an
$\alpha$-value of $0.015$ and a vertical Schmidt number of $2.8$. That fits
almost perfectly in \Fig{f:Sc_alpha}. Since \citet{FromangPapaloizou2006} do
not have an imposed vertical field in their simulations, this may mean that the
rise in Schmidt number with $\alpha$ is something fundamental and not only an
effect of the imposed magnetic field, although further investigations would
have to be done to explore this connection in more detail.

\subsection{Correlation times}

One can express the diffusion coefficient caused by the scale $k$ of a
turbulent flow as $D_k=u_k \ell_k$. Here $u_k$ is the velocity amplitude of
that scale and $\ell_k$ is the typical length-scale over which a turbulent
feature transports before dissolving. The advection length $\ell_k$ can be
approximated by $\ell_k=u_k t_k$, where $t_k$ is the correlation time, or life
time, of a turbulent structure. Taking now an average (and weighted)
correlation time $\tau_{\rm cor}$ of all the scales, one gets the mixing length
expression for the diffusion coefficient in direction $i$,
\begin{equation}
  D_i^{({\rm t})}=\tau_{\rm cor} u_i^2 \, ,
  \label{eq:Dmixing}
\end{equation}
valid for Fickian diffusion \citep[for the validity of Fickian diffusion
see][]{Brandenburg+etal2004}. Here the Mach number, $\sqrt{u_i^2}/c_{\rm s}$, is the
root-mean-square velocity fluctuation in real space. The diffusion
coefficient should thus scale roughly with Mach number squared. We plot the
correlation times, calculated from \Eq{eq:Dmixing}, of $\delta_x$ and $\delta_z$
versus the $\alpha$-value of the flow in \Fig{f:taucor_alpha}.
The correlation time of the turbulent diffusion coefficients falls steeply with
increasing $\alpha$-value, so even though the Mach number of the flow
increases, the time a given turbulent structure has for transporting the dust
becomes shorter and shorter.
Since the correlation times of radial and vertical diffusion have approximately the
same dependence on $\alpha$, the ratio of the diffusion coefficients
can be expressed as $\delta_x/\delta_z = ({\rm
Ma}_x/{\rm Ma}_z)^2$. The anisotropy in the diffusion coefficient in favour of
the radial direction is then mostly an effect of the anisotropy between the
radial and vertical Mach numbers.

\section{Summary}

In this letter we report that the Schmidt number of magnetorotational
turbulence depends strongly on the value of an imposed vertical magnetic
field.  For large values of the vertical field, the relative strength of the
turbulent diffusion falls with respect to the turbulent viscosity. This could
explain part of the discrepancy between measurements of the radial turbulent
diffusion coefficient in magnetorotational without an imposed field
\citep{JohansenKlahr2005} and with an imposed field \citep{Carballido+etal2005}.

\section*{Acknowledgements}

We kindly thank Claire Chandler for encouraging us to conduct the above study.
We would also like to thank Christian Fendt for his critical reading of the
manuscript. Simulations were performed on the PIA cluster of the
Max-Planck-Institut f\"ur Astronomie.

\appendix

\bsp

\label{lastpage}

\end{document}